\documentclass[aps,twocolumn,pra,superscriptaddress,showpacs,tightenlines]{revtex4}
\usepackage{mathrsfs}
\usepackage{amssymb}
\usepackage{amsmath}
\usepackage{graphicx}
\usepackage{epsfig}
\usepackage{txfonts}
\usepackage{subfigure}
\usepackage{amsfonts}
\begin{document}


\title{Quantum speed-up of multiqubit open system via dynamical decoupling pulses }
\author{Ya-Ju Song}
\affiliation{Key Laboratory of Low-Dimensional Quantum Structures
and Quantum Control of Ministry of Education, Department of Physics
and Synergetic Innovation Center for Quantum Effects and
Applications, Hunan Normal University, Changsha 410081, China}
\author{Qing-Shou Tan}
\affiliation{College of Physics and Electronic Engineering, Hainan
Normal University, Haikou 571158, China}
\author{Le-Man Kuang\footnote{Author to whom any correspondence should be
addressed. }\footnote{ Email: lmkuang@hunnu.edu.cn}}
\affiliation{Key Laboratory of Low-Dimensional Quantum Structures
and Quantum Control of Ministry of Education,  Department of Physics
and Synergetic Innovation Center for Quantum Effects and
Applications,  Hunan Normal University, Changsha 410081, China}
\date{\today}

\begin{abstract}
We present a method to accelerate the dynamical evolution of
multiqubit open system  by employing dynamical decoupling pulses
(DDPs) when the qubits are initially in W-type states. It is found
that this speed-up evolution can be achieved in both of the
weak-coupling regime and the  strong-coupling regime. The physical
mechanism behind the acceleration evolution is explained as the
result of the joint action of the non-Markovianity of reservoirs and
the excited-state population of qubits. It is shown that both of the
non-Markovianity and the excited-state population can be controlled
by DDPs to realize the quantum speed-up.

\end{abstract}

\maketitle \narrowtext
\section{\label{Sec:1}Introduction}
Quantum speed limit (QSL) is of particular interest and has
attracted much attention in tremendous areas of quantum physics and
quantum information, such as quantum communication
\cite{Jacob1981,Man2006}, quantum computation
\cite{Warren1993,Lloyd2000,Obada2011}, quantum metrology
\cite{Giovanetti2011,Chin2012,Campo2013,Tsang2013,Alipour2014,Demkowicz2015}
and quantum optical control
\cite{Gordon1997,Rabitz2000,Khaneja2001,Carlini2006,Caneva2009,Mukherjee2013,Hegerfeldt2013,Avinadav2014,Deffner2014}.
In these fields, quantum speed limit time (QSLT) is a key concept,
defined as the minimum evolution time in which a system evolves from
an initial state to a target state. QSLT denoted by $\tau_{QSL}$
determines the theoretical upper bound on the speed of dynamical
evolution. And by first fixing the actual driving time $\tau$, the
shorter QSLT $\tau_{QSL}$, the greater the capacity for potential
speedup will be. In this situation, $\tau_{QSL}=\tau$ means that the
evolution is already along the fastest path and possesses no
potential capacity for further acceleration \cite{Chen2015}.

The speed-up evolution of an open quantum system is of great
significance to the robustness of quantum simulators and computers
against decoherence \cite{Cirac,Geor}.  Since all quantum systems
are unavoidably coupled to their environments, many works have been
done on the derivation of the QSLT for open system dynamics
\cite{Carlini2008,Brody2012,Campo2013,Taddei2013,Sebastian2013,Zhen2014,Zhang2014,Sun2015}
since that for isolated system dynamics
\cite{Mandelstam1945,Uhlmann1992,Pfeifer1993,Margolus1998,Vittorio2003,Chau2010,Deffner2013}.
Generally, these bounds of QSLT could be divided into two
categories:~Mandelstam-Tamm type based on the Cauchy-Schwarz
inequality and Margolus-Levitin type based on the von Neumann trace
inequality. Moreover, Deffner $et$ $al.$ demonstrated that
non-Markovian effects can lead to a smaller QSLT and therefore speed
up quantum evolution \cite{Sebastian2013}. On the other hand, Xu
$et$ $al.$ further proved that the mechanism for speedup is not only
related to non-Markovianity but also to the population of excited
states of the quantum system  under a given driving time
\cite{Zhen2014}. It is the competition between the non-Markovianity
of reservoirs and the excited-state population of qubits that
ultimately determines the acceleration of quantum evolution in
memory environments. A question may now arise how to design a
feasible and effective mechanism to achieve a acceleration process
via modulating the non-Markovianity and the excited-state population
. This question is of particular interest in the weak
system-reservoir coupling regime, where the evolution possesses no
potential capacity to accelerate without any operating on the
system. However, there are few researches on this question
\cite{Zhang2015,Cimmarusti2015}.

In this paper, we propose a new method to accelerate the dynamical
evolution of a multiqubit open system by employing dynamical
decoupling pulses (DDPs). Generally, dynamical decoupling technology
is used to overcome decoherence by averaging the unwanted
interaction with environment to zero
\cite{Lorenza1998,Santos2008,Facchi2004,Rossini2008,Chaudhry2012,Uhrig2007,Pasini2008,
Khodjasteh2007,Jacob2010,Yang2008,Gordon2006,Du2009,Jing2013,Tan2013,Tan2014,Liu}.
Interestingly, we find that, DDPs can also accelerate the dynamical
evolution, or in another word, reduce the time for performing an
elementary logical operation. As a consequence, DDPs can greatly
increase the number of operations within quantum coherence time of
the system. This can help to improve the speed of quantum computers
and communication channels. Meanwhile, compared with a single qubit
system, the multiqubit system has more abundant structures and would
have more extensive applications in quantum information processing.
This is the main motivation of our present work.

In this paper, we consider a multiqubit open quantum system in which
each qubit only interacts with its own reservoir, there is no direct
interaction among qubits, and all reservoirs are independent with
each other. For simplicity, we take the initial states of multiqubit
system under our consideration to be W-type states. We find that the
QSLT has nothing to do with the number of qubits and the
initial-state parameters, but it can be modulated by the number of
DDPs to realize the  acceleration of quantum evolution. And this
speed-up evolution assisted by DDPs can be achieved for both the
weak system-reservoir coupling case and the strong system-reservoir
coupling case. In order to reveal the essential reason of the
acceleration, the roles of DDPs in the non-Markovianity and the
excited-state population are studied in detail.

The paper is organized as follows. In Sec. II, we present the
physical model under our consideration and  derive the dynamics of
multiqubit open system under DDPs. In Sec. III, we investigate the
effect of DDPs on the evolution speed of the multiqubit system and
indicate the speed-up mechanism of the dynamic evolution. Finally,
we conclude our paper with discussions and remarks in the last
section.

\section{\label{Sec:2} Physical Model and solution}

\begin{figure}[tbp]
\includegraphics[bb=-264 267 415 685, width=3.2 in]{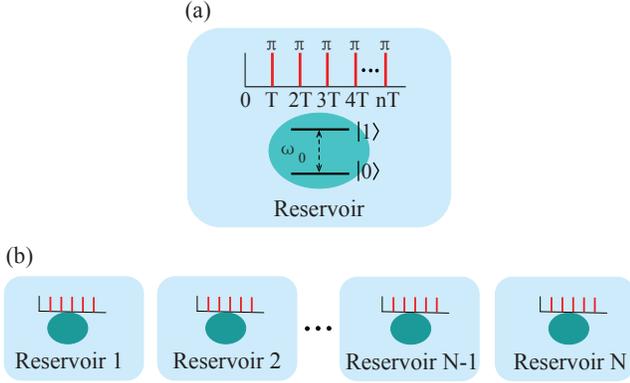}
\caption{(Color online) Schematic diagram of the physical system
considered in this work. (a) A qubit interacts with a reservoir in
the presence of periodic dynamical decoupling pulses (DDPs). The
energy separation of the qubit is $\omega_{0}$, and the periodic
dynamical decoupling is implemented by a train of instantaneous
$\pi$ pulses  applied at the time $T$, $2T$, $3T$, $\cdots$, and
$nT$. (b) $N$ pairs of independent qubit-reservoirs in which DDPs
are applied to each qubit. } \label{fig1.eps}
\end{figure}

We consider $N$ independent qubits interacting with their own
reservoirs locally, and each qubit is controlled by a train of DDPs,
as showed in Fig.~\ref{fig1.eps}. The dynamics of the $N$-qubit open
system can be obtained by the use of time evolution of each
qubit-reservoir pair with the following Hamiltonian \cite{Tan2013}
\begin{eqnarray}
\label{2eq-1}
H &=&\omega _{0}\sigma_{+}\sigma_{-}+\sum_{k}\omega _{k}%
a_{k}^{\dag }a_{k}+\sum_{k}g_{k}\left( \sigma_{+}%
a_{k}+a_{k}^{\dag }\sigma_{-}\right)\nonumber\\
&&+\frac{\pi }{2}\sum_{n=1}^{\infty }\delta \left(t-nT\right)
\sigma_{z},
\end{eqnarray}
where $\omega_{0}$ is the energy difference between the excited
state $|1\rangle$ and the ground state $|0\rangle$ of the qubit,
$\sigma^{\pm }$ are the Pauli raising and lowering operators, and
$a_{k}^\dag (a_{k})$ is the creation (annihilation) operator of the
$k$th mode of the reservoir with the mode frequency $\omega_{k}$.
The coupling strength between the qubit and its reservoir is denoted
by $g_{k}$, which is further characterized by the effective
Lorentzian spectral density \cite{Breuer2002}
\begin{eqnarray}
\label{2eq-2} J(\omega )&=&\frac{1}{2\pi }\frac{\gamma _{0}\lambda
^{2}}{(\omega _{0}-\omega )^{2}+\lambda ^{2}},
\end{eqnarray}
where $\gamma _{0}$ is the Markovian decay rate, $\lambda$ is the
spectral width. Typically, $\gamma_{0}<\lambda/2$ corresponds to the
weak system-reservoir coupling regime, while $\gamma_{0}>\lambda/2$
corresponds to the strong system-reservoir coupling regime. Without
any operation on the system, the dynamics is Markovian in the
weak-coupling regime, while the dynamics  is non-Markovian in the
strong-coupling regime.

The last term in Eq.~(\ref{2eq-1}) corresponds to the dynamical
decoupling control with a train of ideal instantaneous $\pi$ pulses.
$T$ is the time interval between two consecutive pulses. Assuming
that the width of each pulse is sufficiently short, it can be
treated as $\delta$ function. Under the action of each pulse, the
qubit will rotate around the $z$ axis by $\pi$. Then under the
unitary transformation $U\left( t\right)=\hat{\text{T}}\exp [
-i\int_{0}^{t}dt_{1}\frac{\pi }{2}\sum_{n=1}^{\infty }\delta \left(t
_{1}-nT\right) \sigma_{z}]$, Hamiltonian in Eq.~(\ref{2eq-1}) gives
rise to the following effective Hamiltonian
\begin{eqnarray}
\label{2eq-3}
H_{\text{eff}} &=&\omega _{0}\sigma_{+}\sigma_{-}+\sum_{k}\omega _{k}%
a_{k}^{\dag }a_{k}\nonumber\\
&&+\sum_{k}\left( -1\right)
^{n}g_{k}\left( \sigma_{+}a_{k}+a_{k}^{\dag }\sigma%
_{-}\right),
\end{eqnarray}
where $n=\left[ t/T\right] $ is the number of applied pulses within
the driving time $t$, denoted by the integer part of $t/T$. From the
effective Hamiltonian in Eq.~(\ref{2eq-3}), we can see that DDPs
only changes the sign of the qubit-reservoir coupling strength
periodically. This is the physical principle why enough DDPs can
eliminate the effective coupling between the qubit and its
reservoir, and hence the decoherence of open system can be
efficiently suppressed.

Under the single-excitation approximation in a pair of
qubit-reservoir, the reduced density matrix of the qubit at the time
$t$ can be exactly solved as
\begin{eqnarray}
\label{2eq-5'} \rho _{S}\left( t\right) =\left(
\begin{array}{cc}
\rho _{11}\left( 0\right) \kappa _{t}^{2} & \rho _{10}\left(
0\right) \kappa
_{t} \\
\rho _{10}^{\ast }\left( 0\right) \kappa _{t}^{\ast } & 1-\rho
_{11}\left(
0\right) \kappa _{t}^{2}%
\end{array}%
\right),
\end{eqnarray}
or in the form of Kraus operators
\begin{eqnarray}
\label{2eq-4} \rho _{S}(t) &=& \sum_{i=1,2}K_{i}\ \rho_{S} \left(
0\right) K_{i}^{\dag },
\end{eqnarray}
where the two Kraus operators are given by
\begin{subequations}
\label{2eq-5}
\begin{eqnarray}
K_{1} &=&\kappa_{t}\left\vert 1\right\rangle \left\langle
1\right\vert
+\left\vert 0\right\rangle \left\langle 0\right\vert ,\\
K_{2} &=&\sqrt{1-\kappa_{t}^{2}}\left\vert 0\right\rangle
\left\langle 1\right\vert.
\end{eqnarray}
\end{subequations}
When $t\in[ nT,(n+1)T )$, the general solution of $\kappa_{t}$ reads
\cite{Tan2013}
\begin{eqnarray}
\label{2eq-6} \kappa_{t} &=&\left\{
\begin{array}{c}%
e^{-\lambda t/2}\left[ 2\Delta _{n}F_{1n}+\left( 1+\lambda \Delta
_{n}\right) F_{2n}\right],%
\ \ \ \ \ \ \ \ \ \lambda=2 \gamma_{0}, \\%
e^{-\lambda t/2}\left[ A_{n}\cosh \left( \Delta _{n}d\right)
+B_{n}\sinh \left( \Delta _{n}d\right) \right],%
\lambda \neq 2 \gamma_{0},
\end{array}
\right.
\end{eqnarray}
which is a real number. The coefficients on the right-hand side of
Eq.~(\ref{2eq-6}) are given by
\begin{subequations}
\begin{eqnarray}
\label{2eq-7}
F_{1n} = \frac{\lambda ^{2}T\left( \mu_{+}^{n}-\mu_{-}^{n}\right) }{4\sqrt{%
\left( \lambda T\right) ^{2}+4}}\ ,\  F_{2n}
=\frac{\mu_{+}^{n}+\mu_{-}^{n}}{2}+\frac{4 F_{1n}}{\lambda ^{2}T}\ ,\\
A_{n} =\alpha _{+}m_{+}^{n}+\alpha _{-}m_{-}^{n}\ ,%
\  B_{n} =\beta _{+}m_{+}^{n}+\beta _{-}m_{-}^{n}\ ,
\end{eqnarray}
\end{subequations}
where we have introduced the following parameters
\begin{subequations}
\begin{eqnarray}
\label{2eq-8} \Delta _{n} =\frac{t-nT}{2}\ ,\  \mu_{\pm }
=\frac{1}{2}\left[
\lambda T\pm \sqrt{\left( \lambda T\right) ^{2}+4}\right],\\
\alpha _{\pm } =\frac{1}{2}\left[ 1\pm \frac{\cosh \left( \frac{Td}{2}%
\right) }{\xi }\right]\ ,\  %
\beta _{\pm } =\frac{\alpha _{\pm }\left[ m_{\pm }-\cosh \left( \frac{Td}{2%
}\right) \right] }{\sinh \left( \frac{Td}{2}\right) },
\end{eqnarray}
\end{subequations}
with $d=\sqrt{\lambda^{2}-2 \lambda\gamma_{0}}$, $\xi
=\sqrt{1+\left[ \frac{\lambda }{d}\sinh \left( \frac{Td}{2}\right)
\right] ^{2}}$, and $m_{\pm } =\frac{\lambda }{d}\sinh \left(
\frac{Td}{2}\right) \pm \xi $.

From Eq.~(\ref{2eq-5'}), we can find that  the excited-state
population of the qubit at the time $t$ scaled by the excited-state
population of the qubit at the initial time (i.e., the ratio of the
excited-state population of the qubit between at the time $t$ and at
the initial time) can be written as the following simple expression
\begin{eqnarray}
\label{2eq-9} P_{t} &=& \kappa_{t}^{2},
\end{eqnarray}
where $\kappa_{t}$ is given by Eq.~(\ref{2eq-6}). It should be noted
that what Eqs.~(\ref{2eq-5'}) or (\ref{2eq-4}) describe is an
amplitude damping channel model. From Eqs.~(\ref{2eq-5'}) we can see
that $P_{t}$ reflects not only the decay of excited-state
population, but also the decay of quantum coherence of the qubit.

In what follows, we will focus on the situation of $\lambda \neq 2
\gamma_{0}$. In this case,  $P_{t}$ becomes
\begin{eqnarray}
\label{2eq-9'}
P_{t} &=&e^{-\lambda t}\left[ A_{n}\cosh \left( \Delta
_{n}d\right) +B_{n}\sinh \left( \Delta _{n}d\right) \right]^2,
\end{eqnarray}
which indicates that $P_{t}$ can be modulated by the parameters of
coupling spectrum $\gamma_{0}$ and $\lambda$, as well as the number
of DDPs $n$.

Because every pair of qubit-reservoir is independent and identical,
the dynamics of the $N$-qubit open system can be straightforwardly
given by
\begin{eqnarray}
\label{2eq-10} \rho_{t}&=&\sum_{\mu _{1},\ldots \mu _{N}=1,2}[
\otimes _{j=1}^{N}K_{\mu _{j}}(t) ] \rho_{0} [ \otimes
_{j=1}^{N}K_{\mu _{j}}^{\dag }(t)],
\end{eqnarray}
where $K_{\mu _{j}}$ $(\mu_{j}=1,2)$ denotes the Kraus operator for
the $j$th qubit in Eq.~(\ref{2eq-5}). Having obtained the density
operator at any time, one can investigate the QSL of the $N$-qubit
open system as described in the following sections.

\section{\label{Sec:3} Quantum speed-up via DDPs  }

In this section, we study the quantum speed-up of dynamic evolution
of the $N$-qubit open system by employing DDPs. More concretely, we
study the effect of DDPs on the QSLT of this system. This is because
that by first fixing the actual driving time, a decrease in the QSLT
results in an increased ability to speed-up, whereas an increase in
the QSLT means the decrease of the potential speed-up capacity.

Now we begin with the definition of the QSLT in an open system. The
QSLT is defined as the minimum time of  a system evolution from an
initial state $\rho_{0}$ to a target state $\rho_{\tau}$, which is
governed by the time-dependent master equation $ L_{t}\rho
_{t}=\dot{\rho}_{t}$. With the help of the von Neumann trace
inequality and the Cauchy-Schwarz inequality, the QSLT reads
\cite{Sebastian2013}
\begin{eqnarray}
\label{3eq-1}
\tau _{QSL} &=& \frac{\left\vert 1-f_{\tau }\right\vert
}{\min \left\{ E_{1}^{\tau },E_{2}^{\tau },E_{\infty}^{\tau
}\right\} },
\end{eqnarray}
where $E_{p}^{\tau }=\frac{1}{\tau}\int_{0}^{\tau }dt\left\Vert
L_{t}\rho _{t}\right\Vert _{p}$, and $\left\Vert A\right\Vert
_{p}=(\lambda_{1}^{p}+\cdots+\lambda_{n}^{p})^{1/p}$ denotes the
Schatten $p$ norm with descending singular values of operator $A$,
$\lambda _{1}\geq \lambda _{2}\geq \cdots \geq \lambda _{n}$. The
singular values of an operator $A$ are defined as the eigenvalues of
$\sqrt{A^{\dag}A}$. For a Hermitian operator, its singular values
are given by the absolute value of its eigenvalues. And the fidelity
between an initial pure state
$\rho_{0}=|\psi_{0}\rangle\langle\psi_{0}|$ and a target state
$\rho_{\tau}$ is defined as
$f_{\tau}=\langle\psi_{0}|\rho_{\tau}|\psi_{0}\rangle$. Moreover,
the authors in Ref.\cite{Sebastian2013} further proved that the
ML-type bound based on the operator norm ($p=\infty$) provides the
sharpest bound on the QSLT, so we will use the following bound to
derive the QSLT of the $N$-qubit open system,
\begin{eqnarray}
\label{3eq-2} \tau _{QSL}&=&\frac{\tau\left\vert 1-f_{\tau
}\right\vert }{\int_{0}^{\tau }\lambda _{1}dt}.
\end{eqnarray}
Clearly, we can obtain the QSLT by calculating the fidelity
$f_{\tau}$ and the operator norm of $L_{t}\rho_{t}$.

For simplicity, here we assume that the $N$-qubit system is
initially prepared in the W-type states,
\begin{eqnarray}
\label{3eq-3}
\left\vert \psi_{0}\right\rangle &=& \alpha
_{1}\left\vert 100\ldots 0\right\rangle + \alpha _{2}\left\vert
010\ldots 0\right\rangle +\ldots + \alpha _{N}\left\vert 000\ldots
1\right\rangle,\nonumber\\
\end{eqnarray}
where the normalization coefficients satisfy
$\sum_{j=1}^{N}\left\vert \alpha_{j}\right\vert ^{2}=1$. Submitting
$\rho_{0}=|\psi_{0}\rangle\langle\psi_{0}|$ into Eq.~(\ref{2eq-10}),
the reduce density matrix of the $N$-qubit system at the time $t$
reads
\begin{eqnarray}
\label{3eq-4} \rho_{t} &=& P_{t}\left\vert \psi_{0} \right\rangle
\left\langle \psi_{0} \right\vert +\left( 1-P_{t}\right) \left\vert
0\ldots 0\right\rangle \left\langle 0\ldots 0\right\vert,
\end{eqnarray}
where $P_{t}$ is given by Eq.~(\ref{2eq-9'}), and the derivation of
Eq.~(\ref{3eq-4}) is shown in Appendix A. Hence, the fidelity
between the initial W-type states
$\rho_{0}=|\psi_{0}\rangle\langle\psi_{0}|$ and the final state
$\rho_{\tau}$ is found to be
\begin{eqnarray}
\label{3eq-5} f_{\tau } &=& P_{\tau },
\end{eqnarray}
which means that the fidelity between the initial W-type states and
the final state is equal to  the excited-state population in the
final state.

Meanwhile, Eq.~(\ref{3eq-4}) makes
$L_{t}\rho_{t}=\dot{P}_{t}(\left\vert \psi_{0} \right\rangle
\left\langle \psi_{0} \right\vert-\left\vert 0\ldots 0\right\rangle
\left\langle 0\ldots 0\right\vert)$, and
$(L_{t}\rho_{t})^{\dag}=L_{t}\rho_{t}$. Thus, the operator norm of
$L_{t}\rho_{t}$ can be obtained as the absolute value of its
eigenvalues,
\begin{eqnarray}
\label{3eq-6} \lambda _{1} &=& \left\vert \dot{P} _{t}\right\vert.
\end{eqnarray}

Then, using Eq.~(\ref{3eq-2}), one can get the expression of the
QSLT
\begin{eqnarray}
\label{3eq-7} \tau _{QSL} &=& \frac{\tau(1-P_{\tau })}{1-P_{\tau
}+2\int_{0,\dot{P}_{t}>0}^{\tau }\dot{ P}_{t}dt},
\end{eqnarray}
which reflects the speed of the dynamic evolution from the initial
W-type states $|\psi_{0}\rangle$ to the final state $\rho_{\tau}$ by
an actual evolution time $\tau$.

Interestingly, we note that the QSLT has nothing to do with the
number of qubits $N$ and the initial-state-parameter $\alpha_{j}$
\cite{Chen2015}. This is largely because the characteristic of
dynamics with the initial W-type state, as shown in
Eq.~(\ref{3eq-4}). According to Eq.~(\ref{2eq-9'}) and
Eq.~(\ref{3eq-7}), the QSLT could be modulated by the parameters of
coupling spectrum $\gamma_{0}$ and $\lambda$, as well as the number
of DDPs $n$ within the actual evolution time $\tau$. Thus, the
number of DDPs can be used as a control parameter to improve the
speed of evolution. We shall numerically investigate the quantum
speed-up in the following.

In order to observe the speed-up effect induced by DDPs, let's focus
on the QSLT $\tau _{QSL}$ for a fixed actual evolution time of the
system $\tau$. When the actual evolution time $\tau$ is fixed, for
the QSLT ratio $\tau _{QSL}/\tau<1$ the lower ratio $\tau
_{QSL}/\tau$, equivalently, the much shorter $\tau _{QSL}$, the
greater the capacity for potential speedup. And in the case that the
actual evolution time equals the QSLT, i.e., $\tau _{QSL}/\tau=1$,
the evolution is already along the fastest path and there is no
likelihood of further acceleration.

\begin{figure}
\centering
\subfigure{\includegraphics[clip=true,height=4cm,width=7.6cm]{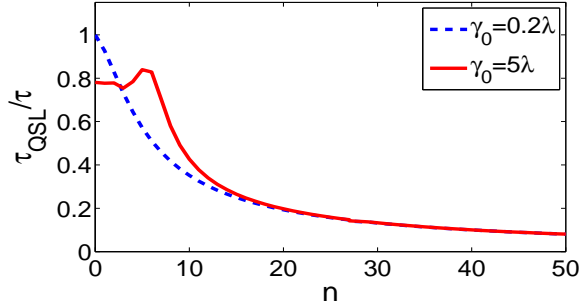}}
\caption{(Color online) The QSLT  versus the number of DDPs $n$ for
a fixed actual driving time. The dash line and the solid line
correspond to $\gamma=0.2\lambda$ (in the weak-coupling regime) and
$\gamma=5\lambda$ (in the strong-coupling regime ), respectively.
Other parameters are chosen as $\lambda\tau=10$ and $\lambda=1$.
}\label{2.eps}
\end{figure}

\begin{figure}
\centering
\subfigure{\includegraphics[clip=true,height=4cm,width=7.6cm]{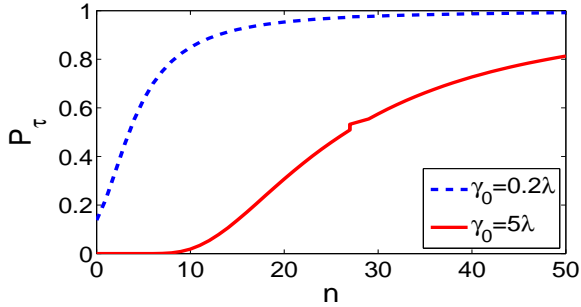}}
\caption{(Color online) The  excited-state population in the final
state $P_{\tau}$ as a function of the number of DDPs $n$. The dash
line and the solid line correspond to $\gamma=0.2\lambda_0$ (in the
weak-coupling regime)
 and  $\gamma=5\lambda_0$ (in the strong-coupling regime ),
respectively. Other parameters are chosen as $\lambda\tau=10$ and
$\lambda=1$. }\label{3.eps}
\end{figure}

In Fig.~\ref{2.eps} we  plot the QSLT  $\tau _{QSL}$ given by
Eq.~(\ref{3eq-7})  versus the number of DDPs $n$ for a fixed actual
driving time $\tau$. From this figure, we can see that in the
weak-coupling regime ($\gamma_{0} =0.2\lambda$, the dash line), the
QSLT  monotonously decreases with the increase of the number of DDPs
$n$. This clearly indicates  the quantum speedup effect induced by
DDPs. However, in the strong-coupling regime ($\gamma_{0}
=5\lambda$, the solid line), the situation is different. From
Fig.~\ref{2.eps} we can see that the QSLT in the strong-coupling
regime first fluctuates in a small $n$ region and then decreases
continuously with the increase of $n$ in the same as the
weak-coupling regime. That is to say, there may exist the speed-down
effect in the strong-coupling regime when the number of DDPs is
small. But anymore, only if the number of DDPs is large enough, DDPs
could induce the shorter QSLT in both the weak-coupling regime and
strong-coupling regime, so as to enhance the capacity for
acceleration. This accelerating effect is especially obvious in the
weak-coupling regime:~in the absence of DDPs (i.e., $n = 0$), the
actual evolution time achieves the QSLT, and there is no likelihood
of acceleration.

\begin{figure}
\centering
\subfigure{\includegraphics[clip=true,height=4cm,width=7.6cm]{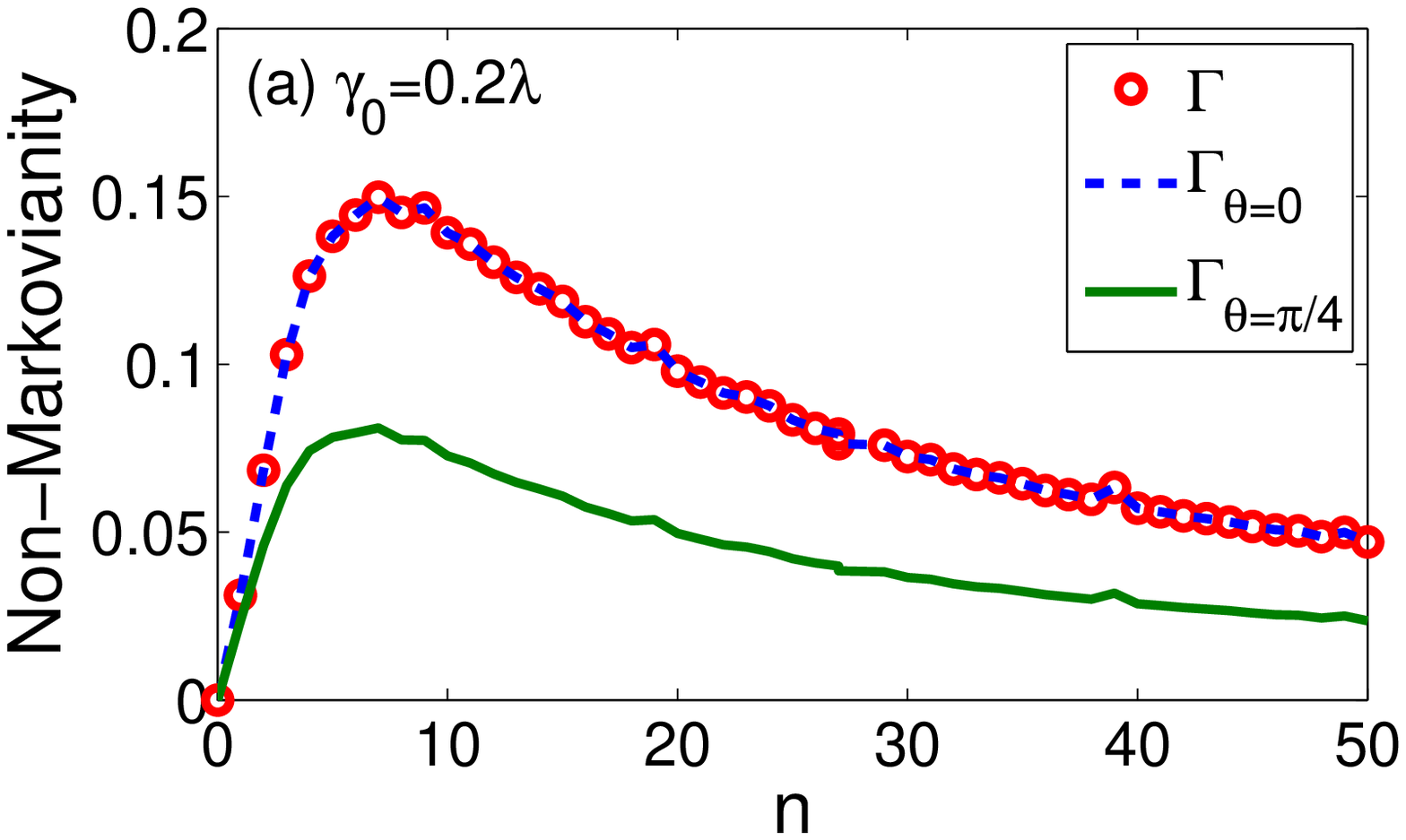}}
\subfigure{\includegraphics[clip=true,height=4cm,width=7.6cm]{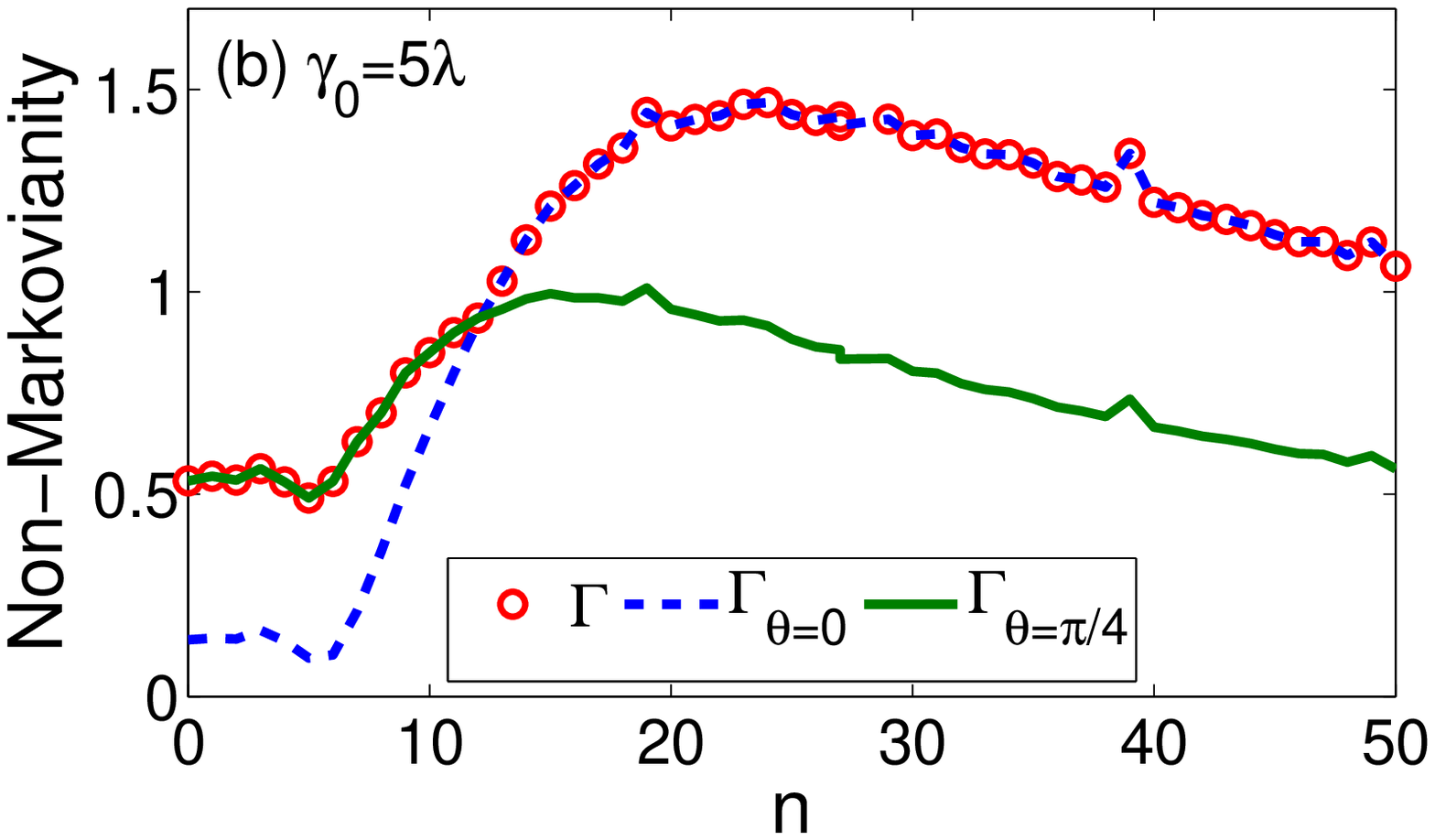}}
\caption{(Color online) The non-Markovianity in a pair of
qubit-reservoir for the fixed driving time $\tau$ as a function of
the number of DDPs $n$. The circle lines represent the
non-Markovianity measure $\Gamma$. The dashed lines and solid lines
correspond  to the non-Markovianity components $\Gamma_{\theta=0}$
and $\Gamma_{\theta=\pi/4}$, respectively. Parameters are chosen as
$\gamma_{0}=0.2\lambda$ for the weak-coupling regime,
$\gamma_{0}=5\lambda$ for the strong-coupling regime,
$\lambda\tau=10$, and $\lambda=1$. }\label{4.eps}
\end{figure}

In order to understand the physical mechanism of the quantum
speed-up assisted by DDPs, we turn to focus on the non-Markovianity
of the reservoir  within the driving time for a pair of
qubit-reservoir \cite{BLP2009} and the excited-state population of
the qubits. Physically speaking, an increase of the
distinguishability of a pair of states during any time interval
implies the emergence of the non-Markovianity. This can be
interpreted as a flow of information from the reservoir back to the
qubit system. In Appendix B, we have derived the non-Markovianity
measure $\Gamma$ characterized in terms of the amount of information
exchanged between the qubit and its reservoir \cite{BLP2009}. By use
of the optimal initial-state pair of the qubit $\left\vert \psi
_{1}\left( 0\right) \right\rangle =\cos \theta \left\vert
1\right\rangle +\sin \theta \exp \left( i\varphi \right) \left\vert
0\right\rangle $ and $\left\vert \psi _{2}\left( 0\right)
\right\rangle = \sin \theta \left\vert 1\right\rangle -\cos \theta
\exp \left( i\varphi \right) \left\vert 0\right\rangle$ with $\theta
\in \left[ 0,\pi/2\right] ,\varphi \in \left[ 0,2\pi \right]$, the
non-Markovianity measure in the time interval $\tau$ can be
expressed as
\begin{eqnarray}
\label{3eq-8}
\Gamma&=&\text{max}\{\Gamma_{\theta=0},\Gamma_{\theta=\pi/4}\},
\end{eqnarray}
where the two non-Markovianity components $\Gamma_{\theta=0}$ and
$\Gamma_{\theta=\pi/4}$ are given by
\begin{eqnarray}
\label{3eq-9}
\Gamma_{\theta=0}=\int_{0,\dot{P}_{t}>0}^{\tau}\dot{P}_{t}dt,\ \
\Gamma_{\theta=\pi/4}=\int_{0,\dot{P}_{t}>0}^{\tau}\frac{\dot{P}_{t}}{2\sqrt{P_{t}}}dt.
\end{eqnarray}
Here the optimal initial-state pair has been numerically proved to
be either $\left\{ \left\vert 0\right\rangle ,\left\vert
1\right\rangle \right\}(\theta =0)$ or $\left\{ \left\vert
+\right\rangle, \left\vert -\right\rangle \right\}(\theta =\pi/4)$
\cite{Steffen2012,Wang2015,He2011,Xu2010}. We can see that the
non-Markovian dynamics appears only when $\dot{P}_{t}>0$, and the
QSLT in Eq.~(\ref{3eq-7}) can be reduced to
\begin{eqnarray}
\label{3eq-10}
\tau_{QSL}&=&\frac{\tau}{1+\frac{2\Gamma_{\theta=0}}{1-P_{\tau}}},
\end{eqnarray}
which indicates that the QSLT is determined by two quantities:~the
non-Markovianity $\Gamma_{\theta=0}$ with the initial-state pair
$\left\{ \left\vert 0\right\rangle ,\left\vert 1\right\rangle
\right\}$ within the evolution time and the excited-state population
in the final state  $P_{\tau}$ in Eq.~(\ref{2eq-9'}). Hence, the
memory effect of reservoir and the excited-state population become
two essential factors for speeding up the dynamical evolution.

As an illustration, we plot the influence of the number of DDPs $n$
on  the excited-state population in the final state  $P_{\tau}$ of
the qubit and the non-Markovianity measure of the reservoir $\Gamma$
in Fig.~\ref{3.eps} and Fig.~\ref{4.eps}, respectively. On one hand,
from Fig.~\ref{3.eps} we can see that in the weak-coupling case (the
dashed line), the value of $P_{\tau}$ of the qubit rises sharply
under the action of DDPs in the small $n$ regime, and it increases
slowly in the large $n$ regime. In particular, $P_{\tau}$ can
approach to unity. These excited-state population changes well match
with the QSLT changes as shown in Fig.~\ref{2.eps}. On the other
hand, from Fig.~\ref{4.eps}(a) we can see that the non-Markovianity
is very weak with the maximum $\Gamma_{max}\simeq 0.15$. This means
the environment of the multiqubit open system under the action of
DDPs is a weak non-Markovian reservoir. Hence, we can conclude that,
although the  non-Markovianity of reservoir is the necessary
condition for speeding up \cite{Chen2015,Zhen2014,Zhang2015}, the
excited-state population is the dominant mechanism of the quantum
speed-up in the weak-coupling case. It is through controlling the
excited-state population of the qubits that DDPs can accelerate the
dynamic evolution of the quantum system under our consideration.

\begin{figure}[tbp]
\centering
\subfigure{\includegraphics[clip=true,height=4.5cm,width=7.6cm]{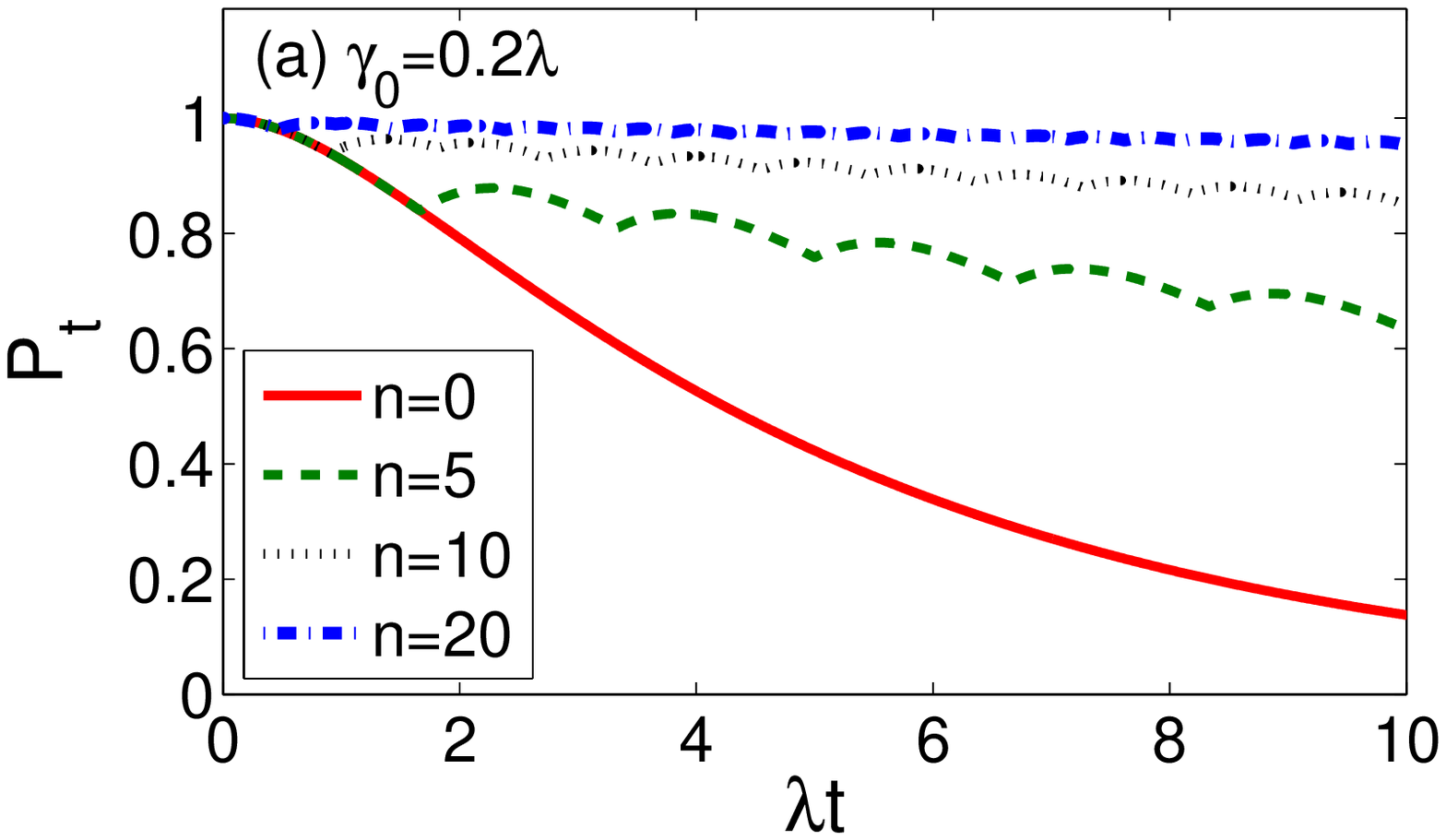}}
\subfigure{\includegraphics[clip=true,height=4.5cm,width=7.6cm]{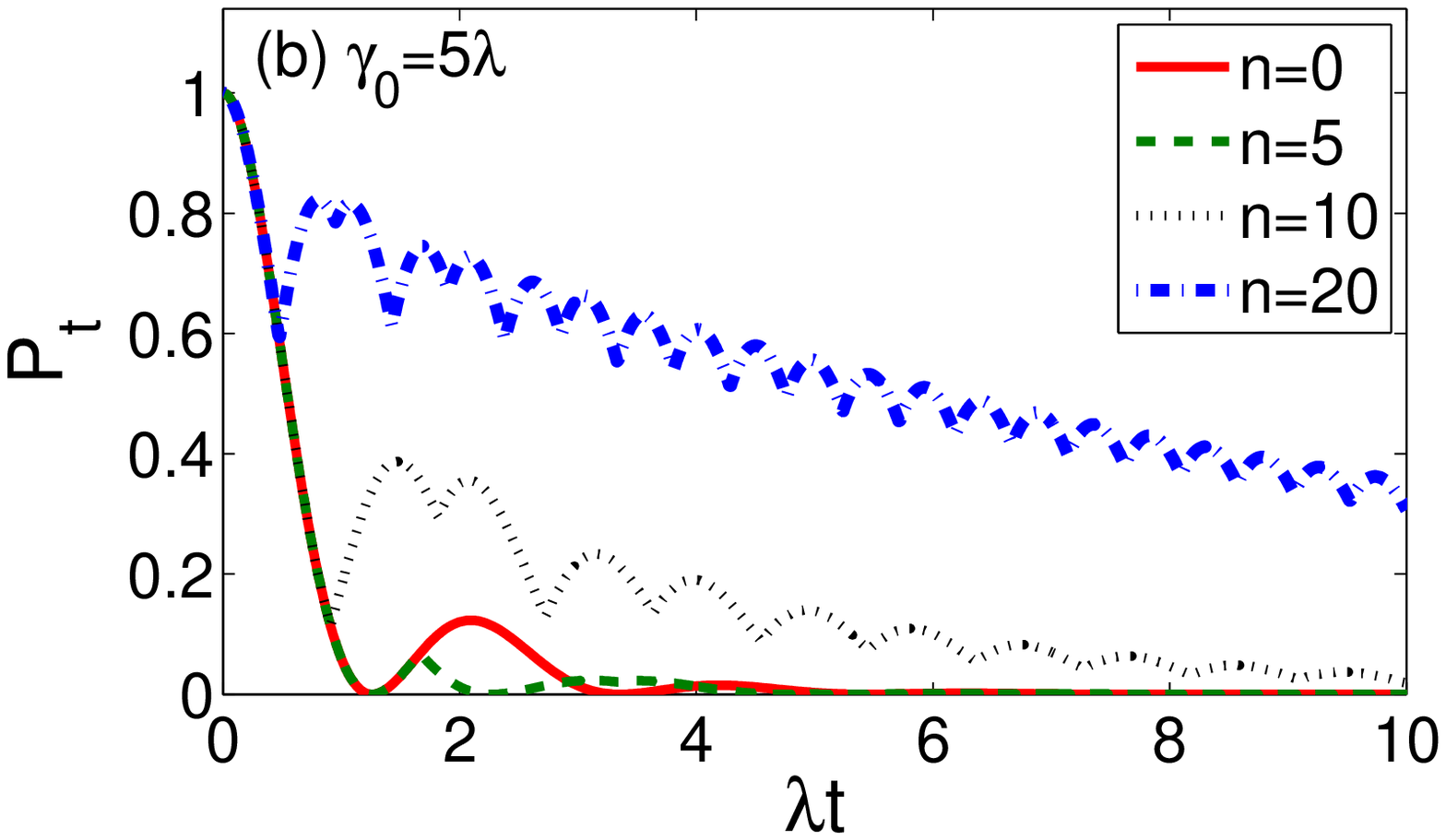}}
\caption{(Color online) The dynamics of $P_{t}$ for different
numbers of DDPs:~$n$=0 (solid lines), $n$=5 (dashed lines), $n$=10
(dot lines), $n$=20 (dot-dashed lines). (a) The case of the
weak-coupling regime $(\gamma_{0}=0.2\lambda)$; (b) The case of  the
strong-coupling regime $(\gamma_{0}=5\lambda)$. $\lambda=1$.}
\label{5.eps}
\end{figure}

In the strong-coupling case (i.e., $\gamma_{0}=5\lambda$), the
influence of  the number of DDPs $n$ on the excited-state population
of the qubit $P_{\tau}$ and the non-Markovianity measure of the
reservoir $\Gamma$ is different from the weak-coupling case. The
solid line in Fig.~\ref{3.eps} reflects changes of the $P_{\tau}$
with respect to the number of DDPs while the circle line in
Fig.~\ref{4.eps}(b) describes changes of the non-Markovianity
measure with respect to the number of DDPs. On one hand, from
Fig.~\ref{3.eps}, we can see that the qubit almost remains in the
zero excited-state population for the small number of DDPs. This
means that the excited-state population does not affect the speed of
the dynamic evolution for the small number of DDPs. On the other
hand, comparing Fig.~\ref{3.eps} with Fig.~\ref{4.eps}(b) for the
strong-coupling case, we can find that the QSLT changing rule under
the small $n$ condition well matches with that of the
non-Markovianity measure. So the non-Markovianity plays the decisive
role on accelerating the dynamic evolution for the small $n$ case in
the strong-coupling regime. From Fig.~\ref{3.eps} and
Fig.~\ref{4.eps}(b), we can see that both of the $P_{\tau}$ and the
non-Markovianity vary soundly with the number of DDPs in the large
$n$ regime of $n>10$. Comparing Fig.~\ref{2.eps} with
Fig.~\ref{3.eps} and Fig.~\ref{4.eps}(b), for the strong-coupling
case we can find that both of the excited-state population and the
non-Markovianity play an important role for the quantum speedup, it
is the competition between the excited-state population and the
non-Markovianity that takes the responsibility to accelerate quantum
evolution in the large $n$ regime of DDPs.

Finally, we discuss the control mechanism of the non-Markovianity of
a reservoir by the use of DDPs. From Fig.~\ref{4.eps} we can see
that on the whole, the non-Markovianity increase firstly and then
decrease later with increasing DDPs number $n$. This can be
explained by the effect of DDPs on the dynamics of each qubit. As
shown in Fig.~\ref{5.eps}, the excited-state population of the qubit
$P_{t}$ fast oscillates as the number of DDPs $n$ increases.
Overall, DDPs not only makes the effective coupling between each
qubit and its reservoir decrease, but also makes the information
exchange between them more rapidly. It is the competition between
these two effects that leads to the nonmonotonic behavior of
$\Gamma$ as a function of $n$. Moreover, contrasting the two
subfigures Fig.~\ref{4.eps}(a) and Fig.~\ref{4.eps}(b), we can see
that the value of $\Gamma$ in strong-coupling regime is greater than
that in weak-coupling regime, while in the weak-coupling regime the
dynamics will be Markovian ($\Gamma=0$) in the absence of DDPs
($n=0$). Besides, the optimal initial-state pairs are also different
in two regimes for calculation of the measure of non-Markovianity.
In the weak-coupling case, the optimal initial-state pair is proved
to be $\{\vert 1\rangle,\vert 0\rangle\}$ ($\Gamma
=\Gamma_{\theta=0}$), while in the strong-coupling case, the optimal
initial-state pair changes from $\{\vert +\rangle,\vert -\rangle\}$
($\Gamma =\Gamma_{\theta=\pi/4}$) to $\{\vert 1\rangle,\vert
0\rangle\}$ ($\Gamma =\Gamma_{\theta=0}$) as $n$ is increased. But
anyway, $\Gamma_{\theta=0}$  reflects the measure of
non-Markovianity.

\section{\label{Sec:4}Conclusions}
In conclusion, we have studied the effects of DDPs on the QSLT of
the $N$-qubit open system when the $N$-qubits are initially in the
 W-type states. We have found that DDPs can be used to
accelerate quantum evolution of multiqubit open system in both of
the weak-coupling and the strong-coupling regimes. In the case of
the weak-coupling between qubits and reservoirs, it has been shown
that the capacity of the quantum acceleration monotonously increases
with the number of DDPs. In the case of the strong-coupling  between
qubits and reservoirs, it is indicated that the quantum speed-up
capacity is the same as that in the weak-coupling case when the
number of DDPs is large enough. While when the number of DDPs is
small, it has been found that quantum speed-down may happen in the
strong-coupling regime between qubits and reservoirs. The essential
physical mechanism for the speed-up evolution is that the
non-Markovianity of reservoirs and the excited-state population of
the qubits jointly determine the QSLT. Under the action of DDPs, the
non-Markovianity of reservoirs and the excited-state population of
the qubits vary so as to lead to the quantum speed-up of the
multiqubit open system. The non-Markovianity of reservoir is the
necessary condition for speeding up in both of the weak- and
strong-coupling regimes. The excited-state population is the
dominant mechanism of the quantum speed-up in the weak-coupling case
while the non-Markovianity of reservoir is the dominant mechanism of
the quantum speed-up in the strong-coupling case when the number of
DDPs is small. Our findings in the present paper may lead to
development of effective methods for the speedup evolution of
multiqubit open systems.  These findings in the present paper could
prove useful to accelerate quantum evolution of multiqubit open
systems. In future work we will explore the possibility of realizing
the quantum speed-up for a more broader class of initial states,
interacted multiqubits, more complicate environments by the use of
DDPs.

\acknowledgments This work was supported by the National Fundamental
Research Program of China (the 973 Program) under Grant No.
2013CB921804 and the National Natural Science Foundation of China
under Grants No. 11375060, No. 11434011 and No. 11447102.

\appendix
\section{Derivation of Eq.~(\ref{3eq-4})}

In this appendix, we present a detailed derivation of quantum state
of the $N$-qubit system at an arbitrary time, i.e., the quantum
state given by Eq.~(\ref{3eq-4}).  When the $N$-qubit system is
initially prepared in the W-type states
\begin{eqnarray}
\label{6eq-1} \left\vert \psi _{0}\right\rangle &=& \alpha
_{1}\left\vert 100\ldots 0\right\rangle +\alpha _{2}\left\vert
010\ldots 0\right\rangle +\ldots +\alpha _{N}\left\vert 000\ldots
1\right\rangle \nonumber\\
&=& \sum^N_{j=1}\alpha _{j}\left\vert j\right\rangle,
\end{eqnarray}
where $\left\vert j\right\rangle $ means that only $j$th qubit is in
the state $\left\vert 1\right\rangle$ and the other $N-1$ qubits are
in the state $\left\vert 0\right\rangle$. Then the density operator
of the initial state can be written as
\begin{eqnarray}
\label{6eq-2} \rho _{0} &=& \left\vert \psi _{0}\right\rangle
\left\langle \psi _{0}\right\vert \nonumber\\
&=& \sum^N_{j=1}\left\vert \alpha _{j}\right\vert ^{2}\left\vert
j\right\rangle \left\langle j\right\vert +\sum_{j\neq k}\alpha
_{j}\alpha _{k}^{\ast }\left\vert j\right\rangle \left\langle
k\right\vert
\end{eqnarray}

For the single qubit case, making use of Eq.~(\ref{2eq-5}) it is
straightforward to show
\begin{subequations}
\label{6eq-3}
\begin{eqnarray}
\sum_{i=1,2}K_{i}\left\vert 0\right\rangle \left\langle
0\right\vert
K_{i}^{\dag } &=&\left\vert 0\right\rangle \left\langle 0\right\vert, \\
\sum_{i=1,2}K_{i}\left\vert 0\right\rangle \left\langle 1\right\vert
K_{i}^{\dag } &=&\kappa _{t}\left\vert 0\right\rangle \left\langle
1\right\vert ,\\
\sum_{i=1,2}K_{i}\left\vert 1\right\rangle \left\langle 0\right\vert
K_{i}^{\dag } &=&\kappa _{t}\left\vert 1\right\rangle \left\langle
0\right\vert, \\
\sum_{i=1,2}K_{i}\left\vert 1\right\rangle \left\langle 1\right\vert
K_{i}^{\dag } &=&\kappa _{t}^{2}\left\vert 1\right\rangle
\left\langle 1\right\vert +\left( 1-\kappa _{t}^{2}\right)
\left\vert 0\right\rangle \left\langle 0\right\vert.
\end{eqnarray}
\end{subequations}

For the $N$-qubit case, by the use of Eq.~(\ref{6eq-3}) we can
obtain the actions of the Kraus operators on  $\left\vert
j\right\rangle \left\langle k\right\vert$ ($j, k=1,2,3, \cdots, N$)
as follows
\begin{widetext}
\begin{subequations}
\label{6eq-4}
\begin{eqnarray}
\sum_{\mu _{1},\mu _{2},\ldots \mu _{N}}\left[ \otimes
_{i=1}^{N}K_{\mu _{i}}\left( t\right) \right] \left\vert
j\right\rangle \left\langle
k\right\vert \left[ \otimes _{i=1}^{N}K_{\mu _{i}}^{\dag }\left( t\right) %
\right] &=& \kappa _{t}^{2}\left\vert j\right\rangle \left\langle k\right\vert,  \hspace{1cm}(j \neq k)\\
\sum_{\mu _{1},\mu _{2},\ldots \mu _{N}}\left[ \otimes
_{i=1}^{N}K_{\mu _{i}}\left( t\right) \right] \left\vert
j\right\rangle \left\langle
j\right\vert \left[ \otimes _{i=1}^{N}K_{\mu _{i}}^{\dag }\left( t\right) %
\right] &=& \kappa _{t}^{2}\left\vert j\right\rangle \left\langle
j\right\vert +\left( 1-\kappa _{t}^{2}\right) \left\vert 0\ldots
0\right\rangle \left\langle 0\ldots 0\right\vert.
\end{eqnarray}
\end{subequations}
\end{widetext}

Making use of Eq.~(\ref{6eq-2}), Eq.~(\ref{6eq-4}) and
Eq.~(\ref{2eq-10}), the reduced density matrix of the $N$-qubit
system at time $t$ can be expressed as
\begin{widetext}
\begin{eqnarray}
\label{6eq-5} \rho _{t} &=&\sum_{\mu _{1},\mu _{2},\ldots \mu
_{N}}\left[ \otimes _{i=1}^{N}K_{\mu _{i}}\left( t\right) \right]
\left[ \sum_{j}\left\vert \alpha _{j}\right\vert ^{2}\left\vert
j\right\rangle \left\langle j\right\vert +\sum_{j\neq k}\alpha
_{j}\alpha _{k}^{\ast }\left\vert j\right\rangle \left\langle
k\right\vert \right] \left[ \otimes
_{i=1}^{N}K_{\mu _{i}}^{\dag }\left( t\right) \right] \nonumber\\
&=&\sum_{j}\left\vert \alpha _{j}\right\vert ^{2}\left[ \kappa
_{t}^{2}\left\vert j\right\rangle \left\langle j\right\vert +\left(
1-\kappa _{t}^{2}\right) \left\vert 0\ldots 0\right\rangle
\left\langle 0\ldots 0\right\vert \right]+\kappa _{t}^{2}\sum_{j\neq
k}\alpha _{j}\alpha
_{k}^{\ast }\left\vert j\right\rangle \left\langle k\right\vert \nonumber\\
&=&\kappa _{t}^{2}\rho _{0}+\left( 1-\kappa _{t}^{2}\right)
\left\vert
0\ldots 0\right\rangle \left\langle 0\ldots 0\right\vert \nonumber\\
&=&P_{t}\left\vert \psi _{0}\right\rangle \left\langle \psi
_{0}\right\vert +\left( 1-P_{t}\right) \left\vert 0\ldots
0\right\rangle \left\langle 0\ldots 0\right\vert,
\end{eqnarray}
\end{widetext}
which exactly is the expression given by Eq.~(\ref{3eq-4}).

\section{The non-Markovianity measure}

The non-Markovianity measure we employ here is based on the amount
of information exchanged between the open system and its reservoir,
which is defined as \cite{BLP2009}
\begin{eqnarray}
\label{5eq-1}
\Gamma&=&\max_{\rho_{1,2}(0)}\int_{\sigma(t)>0}dt\sigma(t).
\end{eqnarray}
where $\sigma (t)$ denotes the rate of change of the trace distance
\begin{eqnarray}
\label{5eq-2}
\sigma(t)&=&\frac{d}{dt}\mathcal{D}\left[\rho_{1}(t),\rho_{2}(t)\right],
\end{eqnarray}
and the trace distance is defined as
\begin{eqnarray}
\label{5eq-3}
\mathcal{D}\left[\rho_{1}(t),\rho_{2}(t)\right]&=&\frac{1}{2}\text{Tr}\left\vert\rho_{1}(t)-\rho_{2}(t)\right\vert.
\end{eqnarray}
Physically, $\sigma (t)<0$ ($\Gamma=0$) holds for all dynamical
semigroups and all time-dependent Markovian processes, while $\sigma
(t)> 0$ ($\Gamma>0$) for non-Markovian dynamics. It is should be
noted that the maximum in Eq.~(\ref{5eq-1}) is taken over all
initial-state pairs of the system. Here we only consider the
non-Markovianity in a pair of qubit-reservoir. Fortunately, it has
been proved that, for the case of a qubit, the optimal initial-state
pairs are always antipodal points on the Bloch sphere
\cite{Steffen2012}. So we can assume $\rho _{1,2}(0) =\left\vert
\psi _{1,2}\left( 0\right) \right\rangle \left\langle \psi
_{1,2}\left( 0\right) \right\vert$, $\left\vert \psi _{1}\left(
0\right) \right\rangle =\cos \theta \left\vert 1\right\rangle +\sin
\theta \exp \left( i\varphi \right) \left\vert 0\right\rangle $ and
$\left\vert \psi _{2}\left( 0\right) \right\rangle = \sin \theta
\left\vert 1\right\rangle -\cos \theta \exp \left( i\varphi \right)
\left\vert 0\right\rangle$ with $\theta \in \left[ 0,\pi/2\right]
,\varphi \in \left[ 0,2\pi \right]$. During the evolution time $t$,
the optimal initial-state pair $\rho _{1}(0)$ and $\rho _{2}(0)$
evolve to the final-state pair $\rho _{1}(t)$ and $\rho _{2}(t)$.
According to Eq.~(\ref{2eq-5'}), we can obtain the simple expression
of the trace distance between two final states at time $t$
\begin{eqnarray}
\label{5eq-4}
\mathcal{D}\left(\rho_{1}(t),\rho _{2}(t)\right)
&=&\sqrt{\cos^{2}(2\theta)P_{t}^{2}+\sin^{2}(2\theta)P_{t}}.
\end{eqnarray}

What is more, as the Bloch sphere governed by Eq.~(\ref{2eq-5'}) is
rotation symmetrical with respect to the $z$ axis, the most
strongest oscillating direction is probably either the pole
direction or any direction in the equatorial plane. In other words,
the optimal initial-state pair should be either $ \left\{ \left\vert
0\right\rangle ,\left\vert 1\right\rangle \right\} (\theta =0)$ or $
\left\{ \left\vert +\right\rangle ,\left\vert -\right\rangle
\right\} (\theta =\pi/4)$ with $\left\vert \pm \right\rangle =\left(
\left\vert 0\right\rangle \pm e^{i\varphi }\left\vert 1\right\rangle
\right) /\sqrt{2}$ \cite{Wang2015,He2011,Xu2010}.

Substituting  Eq.~(\ref{5eq-4}) and Eq.~(\ref{5eq-2}) into
Eq.~(\ref{5eq-1}) we can obtain the simple expression of the measure
of non-Markovianity within the driving time $\tau$
\begin{eqnarray}
\label{5eq-5}
\Gamma&=&\text{max}\{\Gamma_{\theta=0},\Gamma_{\theta=\pi/4}\},
\end{eqnarray}
where the two non-Markovianity parameters  $\Gamma_{\theta=0}$ and
$\Gamma_{\theta=\pi/4}$ are defined by
\begin{subequations}
\begin{eqnarray}
\label{5eq-6}
\Gamma_{\theta=0}&=&\int_{0,\dot{P}_{t}>0}^{\tau}\dot{P}_{t}dt,\\
\Gamma_{\theta=\pi/4}&=&\int_{0,\dot{P}_{t}>0}^{\tau}\frac{\dot{P}_{t}}{2\sqrt{p_{t}}}dt.
\end{eqnarray}
\end{subequations}


\begin{thebibliography}{99}
\bibitem{Jacob1981}      J. D. Bekenstein, Phys. Rev. Lett. \textbf{46}, 623 (1981).
\bibitem{Man2006}        M.-H Yung, Phys. Rev. A. \textbf{74}, 030303(R) (2006).
\bibitem{Warren1993}     W. S. Warren, H. Rabitz, and M. Dahleh, Science \textbf{259}, 1581 (1993).
\bibitem{Lloyd2000}      S. Lloyd, Nature (London) \textbf{406}, 1047 (2000); Phys. Rev. Lett. \textbf{88}, 237901 (2002).
\bibitem{Obada2011}      A.-S. F. Obada, D. A. M. Abo-Kahla, N. Metwally, and M. Abdel-Aty, Physica E \textbf{43}, 1792 (2011).
\bibitem{Giovanetti2011} V. Giovannetti, S. Lloyd, and L. Maccone, Nat. Photonics \textbf{5}, 222 (2011).
\bibitem{Chin2012}       A. W. Chin, S. F. Huelga, and M. B. Plenio, Phys. Rev. Lett. \textbf{109}, 233601 (2012).
\bibitem{Campo2013}      A. del Campo, I. L. Egusquiza, M. B. Plenio, and S. F. Huelga, Phys. Rev. Lett. \textbf{110}, 050403 (2013).
\bibitem{Tsang2013}      M. Tsang, New J. Phys. \textbf{15}, 073005 (2013).
\bibitem{Alipour2014}    S. Alipour, M. Mehboudi, and A. T. Rezakhani, Phys. Rev. Lett. \textbf{112}, 120405 (2014).
\bibitem{Demkowicz2015}  R. Demkowicz-Dobrza$\acute{\mathrm{n}}$ski and M. Markiewicz, Phys. Rev. A. \textbf{91}, 062322 (2015).
\bibitem{Gordon1997}     R. J. Gordon and S. A. Rice, Annu. Rev. Phys. Chem. \textbf{48}, 601 (1997).
\bibitem{Rabitz2000}     H. Rabitz, R. de Vivie-Riedle, M. Motzkus, and K. Kompa, Science \textbf{288}, 824 (2000).
\bibitem{Khaneja2001}    N. Khaneja, R. Brockett, and S. J. Glaser, Phys. Rev. A \textbf{63}, 032308 (2001).
\bibitem{Carlini2006}    A. Carlini, A. Hosoya, T. Koike, and Y. Okudaira, Phys. Rev. Lett. \textbf{96}, 060503 (2006).
\bibitem{Caneva2009}     T. Caneva, M. Murphy, T. Calarco, R. Fazio, S. Montangero, V. Giovannetti, and G. E. Santoro, Phys. Rev. Lett. \textbf{103}, 240501 (2009).
\bibitem{Mukherjee2013}  V. Mukherjee, A. Carlini, A. Mari, T. Caneva, S. Montangero, T. Calarco, R. Fazio, and V. Giovannetti, Phys. Rev. A \textbf{88}, 062326 (2013).
\bibitem{Hegerfeldt2013} G. C. Hegerfeldt, Phys. Rev. Lett. \textbf{111}, 260501 (2013); Phys. Rev. A \textbf{90}, 032110 (2014).
\bibitem{Avinadav2014}   C. Avinadav, R. Fischer, P. London, and D. Gershoni, Phys. Rev. B \textbf{89}, 245311 (2014).
\bibitem{Deffner2014}    S. Deffner, J. Phys. B \textbf{47}, 145502 (2014).
\bibitem{Chen2015}       C. Liu, Z.-Y Xu, and S. Zhu, Phys. Rev. A \textbf{91}, 022102 (2015).

\bibitem{Cirac}          J. I. Cirac and P. Zoller, Nat. Phys. \textbf{8}, 264 (2012).
\bibitem{Geor}           I. M. Georgescu, S. Ashhab, and F. Nori, Rev. Mod. Phys. \textbf{86}, 153 (2014).

\bibitem{Carlini2008}    A. Carlini, A. Hosoya, T. Koike, and Y. Okudaira, J. Phys. A \textbf{41}, 045303 (2008).
\bibitem{Brody2012}      D. C. Brody and E.-M. Graefe, Phys. Rev. Lett. \textbf{109}, 230405 (2012).
\bibitem{Taddei2013}     M. M. Taddei, B. M. Escher, L. Davidovich, and R. L. de Matos Filho, Phys. Rev. Lett. \textbf{110}, 050402 (2013).
\bibitem{Sebastian2013}  S. Deffner and E. Lutz, Phys. Rev. Lett. \textbf{111}, 010402 (2013).
\bibitem{Zhen2014}       Z.-Y. Xu, S. Luo, W.-L. Yang, C. Liu, and S. Zhu, Phys. Rev. A \textbf{89}, 012307 (2014).
\bibitem{Zhang2014}      Y.-J. Zhang, W. Han, Y.-J. Xia, J.-P. Cao, and H. Fan, Scientific reports \textbf{4}, 4890 (2014).
\bibitem{Sun2015}        Z. Sun, J. Liu, J. Ma, and X. Wang, Scientific reports \textbf{5}, 8444 (2015).

\bibitem{Mandelstam1945} L. Mandelstam and I. G. Tamm, J. Phys. (Moscow) \textbf{9}, 249 (1945).
\bibitem{Uhlmann1992}    A. Uhlmann, Phys. Lett. A \textbf{161}, 329 (1992).
\bibitem{Pfeifer1993}    P. Pfeifer, Phys. Rev. Lett. \textbf{70}, 3365 (1993).
\bibitem{Margolus1998}   N. Margolus and L. B. Levitin, Physica D \textbf{120}, 188 (1998).
\bibitem{Vittorio2003}   V. Giovannetti, S. Lloyd, and L. Maccone, Phys. Rev. A. \textbf{67}, 052109 (2003).
\bibitem{Chau2010}       H. F. Chau, Phys. Rev. A. \textbf{81}, 062133 (2010).
\bibitem{Deffner2013}    S. Deffner and E. Lutz, J. Phys. A: Math. Theor. \textbf{46}, 335302 (2013).

\bibitem{Zhang2015}      Y.-J. Zhang, W. Han, Y.-J. Xia, J.-P. Cao, and H. Fan, Phys. Rev. A \textbf{91}, 032112 (2015).
\bibitem{Cimmarusti2015} A. D. Cimmarusti, Z. Yan, B. D. Patterson, L. P. Corcos, L. A. Orozco, and S. Deffner, Phys. Rev. Lett. \textbf{114}, 233602
(2015).

\bibitem{Lorenza1998}    L. Viola and S. Lloyd, Phys. Rev. A \textbf{58}, 2733 (1998); L. Viola, E. Knill, and S. Lloyd, Phys. Rev. Lett. \textbf{82}, 2417 (1999).
\bibitem{Santos2008}     L. F. Santos and L. Viola, New J. Phys. \textbf{10}, 083009 (2008).
\bibitem{Facchi2004}     P. Facchi, D. A. Lidar, and S. Pascazio, Phys. Rev. A \textbf{69}, 032314 (2004).
\bibitem{Rossini2008}    D. Rossini, P. Facchi, R. Fazio, G. Florio, D. A. Lidar, S. Pascazio, F. Plastina, and P. Zanardi, Phys. Rev. A \textbf{77}, 052112 (2008).
\bibitem{Chaudhry2012}   A. Z. Chaudhry and J. Gong, Phys. Rev. A \textbf{86}, 012311 (2012).
\bibitem{Uhrig2007}      G. S. Uhrig, Phys. Rev. Lett. \textbf{98}, 100504 (2007); New J. Phys. \textbf{10}, 083024 (2008).
\bibitem{Pasini2008}     S. Pasini, T. Fischer, P. Karbach, and G. S. Uhrig, Phys. Rev. A \textbf{77}, 032315 (2008).
\bibitem{Khodjasteh2007} K. Khodjasteh and D. A. Lidar, Phys. Rev. Lett. \textbf{95}, 180501 (2005); Phys. Rev. A \textbf{75}, 062310 (2007).
\bibitem{Jacob2010}      J. R. West, D. A. Lidar, B. H. Fong, and M. F. Gyure, Phys. Rev. Lett. \textbf{105}, 230503 (2010).
\bibitem{Yang2008}       W. Yang and R.-B. Liu, Phys. Rev. Lett \textbf{101}, 180403 (2008).
\bibitem{Gordon2006}     G. Gordon and G. Kurizki, Phys. Rev. Lett. \textbf{97}, 110503 (2006); G. Gordon, J. Phys. B \textbf{42}, 223001 (2009).
\bibitem{Du2009}         J. Du, X. Rong, N. Zhao, Y. Wang, J. Yang, and R.-B. Liu, Nature (London) \textbf{461}, 1265 (2009).
\bibitem{Jing2013}       J. Jing, L.-A. Wu, J. Q. You, and T. Yu, Phys. Rev. A \textbf{88}, 022333 (2013).

\bibitem{Tan2013}        Q.-S. Tan, Y. Huang, X. Yin, L.-M. Kuang, and X. Wang, Phys. Rev. A. \textbf{87}, 032102 (2013).
\bibitem{Tan2014}        Q.-S. Tan, Y. Huang, L.-M. Kuang, and X. Wang, Phys. Rev. A. \textbf{89}, 063604 (2014).

\bibitem{Liu}            W. Zhang, J.-L. Hu, J. Zhuang, J. Q. You, and R.-B. Liu. Phys. Rev. B \textbf{82}, 045314 (2010);
                         Z.-Y Wang and R.-B. Liu, Phys. Rev. A \textbf{83}, 022306 (2011);
                         W. Yang, Z.-Y. Wang, and R.-B. Liu, Front. Phys. Chin. \textbf{6}, 2 (2011).

\bibitem{Breuer2002}     H.-P. Breuer and F. Petruccione, \emph{The Theory of Open Quantum Systems} (Oxford University Press, Oxford, 2002).

\bibitem{BLP2009}        H.-P. Breuer, E.-M. Laine, and J. Piilo, Phys. Rev. Lett. \textbf{103}, 210401 (2009); E.-M. Laine, J. Piilo, and H.-P. Breuer, Phys. Rev. A \textbf{81}, 062115 (2010).
\bibitem{Steffen2012}    S. Wissmann, A. Karlsson, E.-M. Laine, J. Piilo,  and H.-P. Breuer, Phys. Rev. A \textbf{86}, 062108 (2012).
\bibitem{Wang2015}       G.-Y. Wang, N. Tang, Y. Liu, and H.-S. Zeng, Chin. Phys. B. \textbf{24(5)}, 050302 (2015);
                         H.-S. Zeng, N. Tang, Y.-P. Zheng, and G.-Y. Wang, Phys. Rev. A \textbf{84}, 032118 (2011);
                         N. Tang, W. Cheng, and H.-S. Zeng, Eur. Phys. J. D \textbf{68}, 278 (2014).

\bibitem{He2011}         Z. He, J. Zou, L. Li, and B. Shao, Phys. Rev. A \textbf{83}, 012108 (2011).
\bibitem{Xu2010}         Z.-Y. Xu, W.-L. Yang, and M. Feng, Phys. Rev. A \textbf{81}, 044105 (2010).

\end{thebibliography}
\end{document}